\documentclass{article}
\usepackage{spconf,amsmath,graphicx}
\usepackage{multirow}
\usepackage{algorithm}
\usepackage[noend]{algpseudocode}
\usepackage{amsmath}
\usepackage{spconf,amsmath,graphicx,xcolor}

\usepackage{amsmath}
\usepackage{mathtools,xparse}
\usepackage{booktabs}
\usepackage{graphics}
\usepackage{multirow}

\title{Unsupervised Neural Mask Estimator For Generalized \\ Eigen-Value Beamforming Based ASR}%
\name{Rohit Kumar, Anirudh Sreeram, Anurenjan Purushothaman, Sriram Ganapathy\thanks{This project was partly funded by grants from Samsung Research India, Bangalore.}}

\address{
 Learning and Extraction of Acoustic Patterns (LEAP) lab, Indian Institute of Science, Bangalore.\\
 Email: \{rohitk, sanirudh, anurenjanr, sriramg\}@iisc.ac.in}
\begin{document}
\ninept
\maketitle
\begin{abstract}
The state-of-art methods for acoustic beamforming in multi-channel ASR are based on a neural mask estimator that predicts the presence of speech and noise. These models are trained using a paired corpus of clean and noisy recordings (teacher model). In this paper, we attempt to move away from the requirements of having supervised clean recordings for training the mask estimator. The models based on signal enhancement and beamforming using multi-channel linear prediction serve as the required mask estimate. In this way, the model training
can also be carried out on real recordings of noisy speech rather than simulated ones alone done in a typical teacher model. Several experiments performed on noisy and reverberant environments in the CHiME-3 corpus as well as the REVERB challenge corpus highlight the effectiveness of the proposed approach. The ASR results for the proposed approach provide performances that are significantly better than a teacher model trained on an out-of-domain dataset  and on par with the oracle mask estimators trained on the in-domain dataset. 
\end{abstract}

\section{Introduction}
\label{sec:intro}
The automatic speech recognition (ASR) in  noisy/reverberant multi-channel environments continue to be a challenging task. The improvement of ASR solutions in such environments are key to several applications like smart speakers, home automation and in meeting transcription systems. The conventional method of processing the multi-channel audio signal involves the spatial filtering performed via beamforming~\cite{van1988beamforming,krim1996two}. The method of beamforming performs a delayed and weighted summation of the multiple spatially separated microphones to provide an enhanced audio signal. The advancements to the basic beamforming using blind reference-channel selection and 
two-step time delay of arrival (TDOA) estimation with Viterbi postprocessing has been proposed to improve the beamforming algorithm~\cite{anguera2007acoustic}. 

An alternate approach to beamforming using a generalized eigen value formulation~\cite{warsitz2007blind} involves a spatial filtering in the complex short-time Fourier transform (STFT) domain. The filter is derived by solving an eigen value problem that maximizes the variance in the ``signal'' direction while minimizing the variance in the ``noise'' direction~\cite{warsitz2007blind} or by keeping the variance in the target direction to be unity while minimizing the variance in the other directions (minimum variance distortionless response (MVDR) beamforming)~\cite{higuchi2016robust}. 

The estimate of speech and noise in the given recording thus becomes the key to perform the beamforming in these approaches. The most successful approach for noise estimation uses a supervised deep neural network (DNN) based speech presence probability (SPP) estimator~\cite{heymann2016neural} at every time-frequency bin. The DNN mask estimator is trained using a pair of clean and multi-channel noisy recordings and the DNN learns the SPP with binary targets. The requirement of parallel speech recordings in clean and multi-channel reverberant conditions is a key limitation to these neural mask estimation methods. Recently, unsupervised approaches to mask estimation using complex mixture Gaussian model have been attempted~\cite{tzinis2019unsupervised,drude2019unsupervised}. However, they are either computationally expensive or suffer from a degradation in performance compared to the DNN based mask estimation using oracle targets.  

In this paper, we propose a combination of multi-channel Linear Prediction (MCLP) based beamforming method~\cite{srikanth} with mask estimation based GEV beamforming for addressing the problem of unsupervised mask estimation and beamforming. The MCLP based algorithm generates a ``clean'' version of the audio that is bootstrapped to a DNN mask estimation process. Using this simple approach, we show that the model can also be effectively trained on real recordings where there are no parallel clean recordings. With several ASR experiments on CHiME-3 and REVERB challenge dataset, we show that the proposed approach performs on par with the oracle mask estimation methods. In addition, the approach significantly improves over a DNN mask estimator trained on an out-of-domain supervised dataset. 

The rest of the paper is organized as follow. Sec.~\ref{sec:gev} describes the DNN mask estimation based GEV beamformer. The proposed unsupervised model for beamforming is given in Sec.~\ref{sec:unsupGEV}. The experiments and results on ASR tasks are reported in  Sec.~\ref{sec:expt}. A summary of the work is given in Sec.~\ref{sec:summary}. 

\begin{figure*}
  \centering
  \includegraphics[width=\textwidth,height=5.7cm]{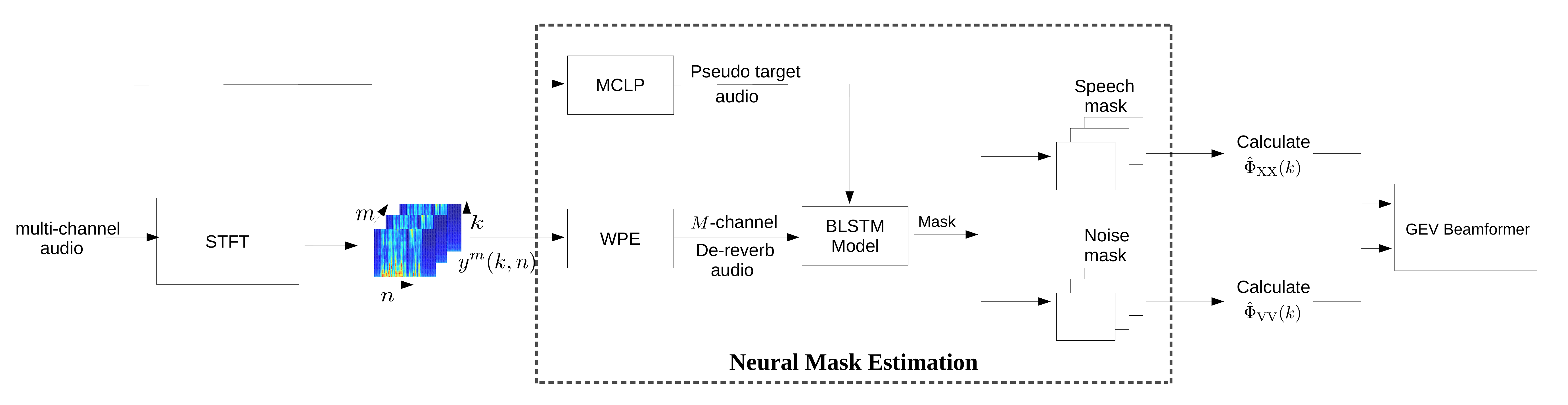}
  \vspace{-0.6cm}
      \caption{Block schematic of the unsupervised neural network based mask estimation.}
      \vspace{-0.2cm}
  \label{fig2}
\end{figure*}
\section{DNN mask estimation based GEV beamformer}\label{sec:gev}
\subsection{Speech Pre-processing}
Let the  observed speech signal in $m^{th}$ microphone be represented by $y^m(k,n)$ in the short time Fourier transform (STFT) domain, where $k$ denotes the frequency bin index and $n$ denotes the frame index. This observed signal is corrupted with reverberation  and additive noise, $v^m(k,n)$. 
\begin{equation}\label{eq:sig_model}
y^m(k,n) = \sum_{l=0}^{L_h-1}g^m(k,l)x(k,n-l) + v^m(k,n)
\end{equation}
where $g^m(k,n)$ is the STFT of the room response function and $x(k,n)$ is the source signal STFT.  

\subsection{Generalized Eigen Value (GEV) Beamforming}
The beamforming operation in frequency domain determines the spatial filter coefficients $w(m,k)$ to obtain the enhanced signal,
\begin{equation}\label{eq:beamformer}
z(k,n) = \sum _{m=0}^{M-1} w(m,k)~ y^m(k,n)    
\end{equation}
where $z(k,n)$ is the beamformed signal. 
The main goal of GEV beamforming is to determine the spatial filter coefficients  $\textbf{w}(k) = [w(0,k),..,w(M-1,k)]^T$ such that the SNR at the output of the filter is maximized~\cite{warsitz2007blind}, i.e., 
\begin{equation}\label{eq:gev}
    \textbf{w}_{GEV}(k) = \arg\max_{\textbf{w}(k)} \frac{\textbf{w}^{H}(k) \boldsymbol {\hat{\Phi}}_{XX}(k)\textbf{w}(k)}{\textbf{w}^{H}(k) \boldsymbol {\hat{\Phi}}_{VV}(k) ~\textbf{w}(k)}
\end{equation}
\newline where $\boldsymbol {\hat{\Phi}}_{XX}$ and $\boldsymbol {\hat{\Phi}}_{VV}$ are power spectral density (PSD) estimates of the clean signal and noise respectively.

The most successful approach to the estimation of clean and noise PSD is through the use of a neural mask estimator (described next). Once the PSD matrices are estimated, the solution to the optimization given in Eq.~(\ref{eq:gev}) is the eigen vector corresponding to maximum eigen value of the matrix   $\boldsymbol {\hat{\Phi}}_{VV}^{-1} \boldsymbol {\hat{\Phi}}_{XX}$.
\subsection{MVDR Beamforming}
The most commonly used beamforming method is MVDR based beamforming. This formulation tries to minimize the residual noise keeping the constraint that the signal from preferred source direction being distortionless,
\begin{equation}
    \textbf{w}_{MVDR}(k) = \frac{ \boldsymbol {\hat{\Phi}}^{-1}_{VV}(k)\textbf{d}}{\textbf{d}^{H} \boldsymbol {\hat{\Phi}}^{-1}_{VV}(k) ~\textbf{d}}
\end{equation}
where $\textbf{d}$ specifies the preferred direction of arrival.

\subsection{Neural Mask Estimator}
As proposed in \cite{heymann2016neural,nakatani2017integrating}, the neural mask estimators are deep feed-forward/recurrent networks that are trained to predict the speech presence probability in each time-frequency bin. In simulated settings (where $y^m(k,n)$ and $x(k,n)$ are available), the deep model is trained with magnitude STFT $|y^m(k,n)|$ coefficients for patch of frames $n$ and all frequency bins to predict the ideal ratio mask (IRM). The IRM is obtained by thresholding the ratio of magnitude STFT $\frac{|y^m(k,n)|}{|x(k,n)|}$  with a threshold different for voiced and unvoiced regions of the audio~\cite{heymann2016neural}.  The output of the mask estimator performs a sigmoid non-linearity and these outputs are interpreted as speech presence probability estimators $s(k,n)$. Once the mask estimator is trained, the PSD matrices needed in Eq.~(\ref{eq:gev}) for $\textbf{y}(k,n) = [y^0(k,n),..,y^{M-1}(k,n)]^T$ is, 
\begin{eqnarray}\label{eq:psd_estimate}
\boldsymbol {\hat{\Phi}}_{XX}(k) = \frac {\sum _n s(k,n) \textbf{y}(k,n) (\textbf{y}(k,n))^H}{\sum _n s(k,n)} 
\end{eqnarray}
The noise PSD is also estimated in a similar fashion using $ (1 - s(k,n))$ as the mask. 

\section{Unsupervised Mask Estimation}\label{sec:unsupGEV}
One of the limitations of the neural mask estimation described above is the need for simulated data with parallel clean and noisy multi-channel recordings to train the deep model. Hence, the real multi-channel recordings cannot be used in the neural mask training. In this paper, we propose to move away from the requirement of having simulated settings by generating unsupervised pseudo targets for the real (and simulated) multi-channel recordings. The block schematic of the unsupervised mask estimation algorithm is given in Fig.~\ref{fig2}.

  
\subsection{Target computation for mask estimation}
We use the joint spatial filtering and multi-channel linear prediction (MCLP) approach with Bayesian inference proposed in \cite{srikanth,chetupalli2019late} to derive the unsupervised targets for the neural mask estimator. For a single reference signal characterized by STFT coefficients $d_1(k,n)$, the MCLP model for a $m^{th}$ microphone signal ~\cite{srikanth} is given as,
\begin{equation}
y^m(k,n) = a^m(k) d_1(k,n) + (\textbf{g}^m(k))^H\boldsymbol{\phi}(k,n)
\end{equation}
where $y^m(k,n)$ is the STFT of the $m$th microphone signal, $a^m[k]$ is the relative gain of the desired signal collected at the $m$th microphone. The late reflection components in the multi-channel signal are modeled as a linear prediction with $\textbf{g}^m(k)$ denoting a vector of $LM$ prediction coefficients and $\boldsymbol{\phi}(k,n) = [y^1(k,n-D-1),..,y^1(k,n-D-L),...,y^M(k,n-D-1),..,y^M(k,n-D-L)]^T$ is the $LM$ dimensional vector containing the delayed STFT components from all the $M$ microphones for $L$ previous lags. In vector form, 
\begin{equation}
    {\bf y}(k,n) = {\bf a}(k) d_1(k,n) + {\bf G}(k)^H \boldsymbol{\phi}(k,n)
\end{equation}
where ${\bf G}[k]$ is the $M \times LM$ MCLP filter coefficients, ${\bf a}[k]$ is the relative transfer function (RTF) of each microphone with respect to the reference signal. 

A spatial filter $\textbf{w}(k)$ is constructed such that $\textbf{w}^H(k) \textbf{a}(k) = 1$. This gives,
\begin{equation}
    \textbf{w}^H(k)  {\bf y}(k,n) = d_1(k,n) + \textbf{w}^H(k){\bf G}(k)^H \boldsymbol{\phi}(k,n)
\end{equation}
By assuming a complex circular Gaussian prior on the desired source signal, $p(d_{1}(k,n)) \sim N_{c}(d_{m}(k,n);0,\gamma_{kn})$, a maximum likelihood (ML) approach to parameter estimation can be pursued~\cite{chetupalli2019late}. This ML problem can be solved using a coordinate ascent method where the parameters of the model ($\Theta$) containing the MCLP prediction coefficients ${\bf G}(k)$, the RTF ${\bf a}(k)$, the spatial filter $\textbf{w}(k)$ and the unknown variance $\gamma_{kn}$ are iteratively estimated. 

The solution to the ML estimation problem~\cite{chetupalli2019late} is given below. The ML problem can be equivalently stated as,
\begin{multline}
    maximize~~\sum_{n=0}^{N-1} \gamma_{kn}^{-1} \left|{\bf w}[k]^H \left[ {\bf y}[k,n] - {\bf G}[k]^H {\boldsymbol \phi}[k,n] \right] \right|^2 ,\\~subject~to~{\bf w}[k]^H {\bf a}[k]=1.
\end{multline}

The prediction filter ${\bf G}[k]$, spatial filter ${\bf w}[k]$ and the RTF ${\bf a}[k]$ are estimated sequentially in an iterative scheme, using the equations given below
\begin{equation}
    \hat{\bf G}(k)=\textbf{R}_{\boldsymbol{\phi \phi}}^{-1}(k)\textbf{R}_{\boldsymbol {\phi{y}}}(k)
\end{equation}
where
\begin{eqnarray}
   \textbf{ R}_{\boldsymbol {\phi \phi}}(k)=\sum_{n=0}^{N-1}{\gamma_{kn}^{-1}}{\boldsymbol \phi}(k,n)\boldsymbol{\phi}^H(k,n) \\
   \textbf{ R}_{\boldsymbol{\phi{y}}}(k)=\sum_{n=0}^{N-1}{\gamma_{kn}^{-1}}{\boldsymbol \phi}(k,n)\textbf{y}^H(k,n)
\end{eqnarray}
\begin{table}[t!]
\caption{CLSTM model architecture.}
\vspace{+0.2cm}
\centering
\begin{tabular}{c|c}
\hline
\textbf{Layer} & \textbf{Configuration}      \\ \hline
Conv 2D (ReLU) & filters = 128, kernel (3,3) \\
Conv 2D (ReLU) & filters = 128, kernel (3,3) \\
Maxpooling 2D  & size = (2,2)                \\
Conv 2D (ReLU) & filters = 64, kernel (3,3)  \\
Conv 2D (ReLU) & filters = 64, kernel (3,3)  \\
LSTM (ReLU)    & 1024 units (frequency recurrence)                  \\
DNN (ReLU)     & 1024 units                  \\
DNN (ReLU)     & 1024 units                  \\
DNN (Softmax)  & senone posteriors          \end{tabular}
\label{table:1}
\vspace{-0.3cm}
\end{table}

Once the MCLP prediction coefficients $ \hat{\bf G}(k)$ are estimated, the RTF vector $\textbf{a}(k)$ can estimated as the first column of the prediction residual, i.e., ${\bf y}[k,n] - \hat{{\bf G}}[k]^H {\boldsymbol \phi}[k,n]$. 

Let $\textbf{R}_{\hat{r}\hat{r}}$ denote the spatial correlation matrix of the predicted reverberation component $\hat{\textbf{r}}=\hat{\textbf{G}}^{H}(k){\boldsymbol \phi}(k,n)$. Then, the spatial filter can be estimated as,
\begin{equation}
    \hat{\textbf{w}}(k) = \frac{\textbf{R}_{\hat{\textbf{r}}\hat{\textbf{r}}}^{-1}\hat{\textbf{a}}}{\hat{\textbf{a}}^{H}\textbf{R}_{\hat{\textbf{r}} \hat{\textbf{r}}}^{-1}\hat{\textbf{a}}}.
\end{equation}
Finally, the desired signal variance $\gamma_{kn}$ is estimated using an AR modeling approach on the estimate of the desired signal $d_1(k,n)~$\cite{srikanth}. More details on the ML estimation can be found in~\cite{chetupalli2019late,srikanth}

Using the iterative procedure outlined above, the estimation of the late reflection components and beamforming of the desired source signal are jointly performed. The output estimate of $d_1(k,n)$ is used as the estimate of the clean signal in the GEV beamforming (Fig.~\ref{fig2}). 

\section{Experiments and Results}\label{sec:expt}
\subsection{Mask Estimation}
The experimental setup for all the experiments are as follows. A $512$ point Short Time Fourier Transform (STFT) of the multi-channel audio signal is computed jointly to form a $3$ dimensional ($F$$\times$$T$$\times$$M$) tensor where length, height and width represents number of frequency bins $F$, time frames $T$ and number of channels $M$ respectively. By segregating voiced and unvoiced section in each frequency bin, an ideal ratio mask (IRM) is estimated for the 3D input using the MCLP based beamformed target~\cite{heymann2016neural}. 

The model architecture for the mask estimation uses a Bi-directional Long Short-term memory (BLSTM) network followed by two fully connected layers. We use Rectified linear unit (ReLU) activation function for the first two layers and Sigmoid for the last layer. A dropout regularization is used with dropout parameter of $0.5$ after every layer. For training the unsupervised model, the targets are derived from the audio beamformed using the method of multi-channel linear prediction described in Sec.~\ref{sec:unsupGEV}. The speech and noise masks are estimated using the model for all the channels jointly. A single speech mask and noise mask (complimentary to the speech mask) are generated by taking the median of all the masks from the multiple channels.  The $\boldsymbol{\hat{\Phi}}_{XX}$ and $\boldsymbol{\hat{\Phi}}_{VV}$ are calculatedand the beamformed STFT  estimate is then converted back to the audio signal using overlap synthesis. These audio signals are converted to acoustic features for ASR training and testing. 
 \begin{table}[t!]
\caption{Word Error Rate (\%) for CHiME-3 dataset.}

\resizebox{\columnwidth}{!}{
\begin{tabular}{@{}l|ccc|ccc@{}}
\toprule
\multicolumn{1}{c|}{\multirow{2}{*}{\begin{tabular}[c]{@{}c@{}} Training  \end{tabular}}}                        & \multicolumn{3}{c|}{Dev} & \multicolumn{3}{c}{Eval} \\ \cmidrule(l){2-7} 
\multicolumn{1}{c|}{}                                                    & Real    & Sim & Avg   & Real    & Sim   & Avg   \\ \midrule
\begin{tabular}[c]{@{}l@{}}BeamformIt~\cite{anguera2007acoustic} \end{tabular} &6.1      &8.4     &7.3  &13.0 &12.7   &12.9  \\
\begin{tabular}[c]{@{}l@{}}3-D CNN~\cite{ganapathy} \end{tabular} &7.2      &7.2     &7.2      &15.4      &9.1   &12.2  \\

\begin{tabular}[c]{@{}l@{}}Sup. Out-of-dom. GEV  \end{tabular} &7.1      &{8.8}     & {7.9}     &{11.2}     &{10.7}   & {10.9} \\ 
\begin{tabular}[c]{@{}l@{}}Unsup. MVDR \end{tabular} &4.9      &6.2     &5.5      &9.4     &7.4   &8.4 \\
\begin{tabular}[c]{@{}l@{}}Unsup. GEV \end{tabular} &{\textbf{4.9}}      &{\textbf{5.8}}     & {\textbf{5.3}}     &\textbf{9.0}     &{\textbf{7.3}}   &\textbf{ 8.1 }\\ \hline \hline
\begin{tabular}[c]{@{}l@{}}Sup. oracle MVDR~\cite{nakatani2017integrating} \end{tabular} &5.1      &6.5     &5.8     &9.1     &7.5   &8.3  \\
\begin{tabular}[c]{@{}l@{}}Sup. oracle GEV~\cite{heymann2016neural} \end{tabular} &4.9      &{6.1}     & {5.5}     &{9.4}     &7.2   & {8.3} \\ \hline
\end{tabular}
\vspace{-2.5cm}
\label{table:2}}
\end{table}

\subsection{ASR setup}
The ASR system uses filter-bank (FBANK) features that are $40$ log-mel spectrogram features extracted  every $25$ms  windows  with  a  shift  of $10$ms on multi-channel audio signals that are enhanced with WPE~\cite{wpe}. We use the Kaldi toolkit~\cite{povey2011kaldi} for deriving the senone alignments used  in  the  PyTorch  deep  learning  framework.    A  hidden Markov  model  -  Gaussian  mixture  model  (HMM-GMM) system is initially trained  to generate the alignments. The acoustic model used in this work is a convolutional long short term memory (CLSTM) model where the LSTM recurs over frequency. The configuration of the CLSTM model is given in Table~\ref{table:1}. A dropout of $20\%$ and batch normalization is used after every layer for regularization. For the ASR decoding, an initial tri-gram model is used to generate a lattice rescored with a recurrent neural network   (RNN)~\cite{mikolov2010recurrent}.

The proposed method of beamforming using the psuedo mask estimates from a multi-channel linear prediction based beamformer is compared with the beamforming using delay-sum and Viterbi algorithm (BeamformIt~\cite{anguera2007acoustic}), a 3-D CNN based neural acoustic model which jointly performs beamforming and ASR~\cite{ganapathy} and the generalized eigen-value (GEV) based beamforming with supervised mask estimation on the simulated data~\cite{heymann2016neural}. 

\subsection{CHiME-3 ASR}
The CHiME-3 corpus for ASR contains multi-microphone tablet device recordings from everyday environments, released as a part of 3rd CHiME challenge \cite{chime3}. Four varied environments are present, cafe (CAF), street junction (STR), public transport (BUS) and pedestrian area (PED). For each environment, two types of noisy speech data are present, real and simulated. The real data consists of $6$-channel recordings of sentences from the WSJ$0$ corpus spoken in the environments listed above. The simulated data was constructed by artificially mixing clean utterances with environment noises. The training data has $1600$ (real) noisy recordings and $7138$ simulated noisy utterances.  The development (dev) and evaluation (eval) data consists of the $410$ and $330$ utterances respectively. For each set, the sentences are read by four different talkers in the four CHiME-3 environments. This results in $1640$ ($410 \times 4$) and $1320$ ($330 \times 4$) real development and evaluation utterances in total. Identically-sized, simulated dev and eval sets are made by mixing recordings captured in the recording booth with the environmental noise recordings.

\begin{table}[t]
\begin{center}
\caption{WER (\%) for each noise condition in CHiME-3 dataset with supervised and unsupervised GEV beamforming methods.}
\label{tab:Chime3Results_detailed_rnnlm}
    \resizebox{8.65cm}{1.2cm}{
	\begin{tabular}{|c|c|c|c|c||c|c|c|c|}
	\hline
	 & \multicolumn{4}{c||}{\textbf{Dev Data}} & \multicolumn{4}{c|}{\textbf{Eval Data}} \\\hline
	\multirow{2}{*}{Cond.} & \multicolumn{2}{c|}{Sim} & \multicolumn{2}{c||}{Real} & \multicolumn{2}{c|}{Sim} & \multicolumn{2}{c|}{Real} \\\cline{2-9}
	& Sup & Unsup & Sup & Unsup & Sup & Unsup & Sup & Unsup \\\hline
	BUS &4.9  &4.9  &6.0  &6.0  &5.9  &6.1  &12.8   &11.9 \\
	CAF &8.1 &7.6  &4.8  &4.8  &8.3  &8.1  &8.8  &8.5  \\
	PED &5.6  &5.3  &4.2 &4.2  &7.1  &7.0  &9.2  &8.9 \\
	STR &5.7  &5.4  &4.6  &4.7 &7.3  &7.9  &6.8  &6.6  \\ \hline
	\end{tabular}
	}
	\vspace{-0.7cm}
\end{center}
\label{table:3}

\end{table}
 
The results for the CHiME-3 ASR system with various beamforming methods are given in Table~\ref{table:2}. The ASR results for the BeamformIt~\cite{anguera2007acoustic} are similar to the 3-D CNN model~\cite{ganapathy}. The GEV based on out-of-domain set consists of training the neural mask estimation on Reverb Challenge dataset (described next) and using the mask estimator outputs for GEV beamforming on CHiME-3 dataset. While there is a domain mis-match, this approach provides the best baseline beamforming system, particularly on the evaluation data of CHiME-3. 
The supervised GEV/MVDR using oracle mask estimates on the in-domain CHiME-3 dataset provides the upper bound in terms of the performance of the unsupervised methods. The proposed unsupervised GEV beamforming using the MCLP based source signal targets provides very similar results to the supervised oracle mask estimation based GEV. In the case of real recordings in the evaluation data, the unsupervised method has a superior performance compared to the supervised method as the supervised method of mask estimation uses only the simulated data. In terms of relative improvements over the BeamformIt method and the out-of-domain mask estimation based GEV, the proposed approach yields about $27$ \% and $35$ \% respectively on the development data and $37$ \% and $25$ \% respectively on the evaluation data. 

The comparison of the supervised and unsupervised approaches on the different noise conditions of the CHiME-3 dataset are shown in Table~\ref{table:3}. As seen in this Table, for most of the noise conditions, the unsupervised method compares well with the supervised mask estimation approach. A degradation is seen in the unsupervised case for ``Street'' noise in simulated conditions. However, a good improvement in ASR performance is seen for ``Bus'' noise in real evaluation conditions for the unsupervised approach as well.

\subsection{Reverb Challenge}
The Reverb Challenge dataset \cite{kinoshita2013reverb} contains recordings with real and simulated reverberation condition, recorded using 8 channels for the ASR task. The  simulated  data  is  comprised  of  reverberant  utterances  generated  (from  the  WSJCAM0 corpus)  obtained  by  artificially convolving  clean  WSJCAM0  recordings  with  the  measured  room impulse responses (RIRs) and adding noise at an SNR of 20 dB. The real data consists of utterances spoken by human speakers in a noisy reverberant room, with utterances from the multi-channel Wall Street Journal audio-visual (MC-WSJ-AV) corpus [3]. The training set consists of 7861 utterances (92 speakers) from the clean WSJCAM0 training data by convolving the clean utterances with 24 measured RIRs.  The development (Dev.) and evaluation (Eval.) datasets consists of 1663 (1484 simulated  and 179 real) recordings and 2548 (2176 simulated and 372 real) recordings respectively. The Dev. and Eval. datasets have 20 and 28 speakers respectively.

The ASR results for the various beamforming methods on the Reverb Challenge dataset are shown in Table~\ref{table:4}. The unsupervised beamforming method improves significantly over the BeamformIt method and the 3-D CNN approach. On the average, the unsupervised mask estimation approach performs similar to the supervised mask estimation approach in the GEV/MVDR beamforming. The unsupervised approach improves the BeamformIt approach relatively by $18$\% on the development data and $35$ \% on the evaluation data. The ASR results on the Reverb Challenge dataset are also seen to be consistent with those for the CHiME-3 dataset.
\begin{table}[t!]
\caption{Word Error Rate (\%) for REVERB Challenge dataset using various beamforming methods.}

\resizebox{\columnwidth}{!}{
\begin{tabular}{@{}l|ccc|ccc@{}}
\toprule
\multicolumn{1}{c|}{\multirow{2}{*}{\begin{tabular}[c]{@{}c@{}} Training  \end{tabular}}}                        & \multicolumn{3}{c|}{Dev} & \multicolumn{3}{c}{Eval} \\ \cmidrule(l){2-7} 
\multicolumn{1}{c|}{}                                                    & Real    & Simu   & Avg   & Real    & Simu   & Avg   \\ \midrule
\begin{tabular}[c]{@{}l@{}}BeamformIt~\cite{anguera2007acoustic} \end{tabular} &19.7      &6.2     &12.9  &22.2 &6.5   &14.4  \\
\begin{tabular}[c]{@{}l@{}}3-D CNN~\cite{ganapathy} \end{tabular} &20.4     &6.7     &13.5      &21.2      &6.6   &13.9  \\
\begin{tabular}[c]{@{}l@{}}Unsup. MVDR\end{tabular} &17.2      &\textbf{5.1}     &11.2      &14.9    & 5.6  &10.3 \\
\begin{tabular}[c]{@{}l@{}}Unsup. GEV\end{tabular} &\textbf{15.6}      &5.6     & \textbf{10.6}     &{\textbf{13.5}}     &{\textbf{5.3}}   & {\textbf{9.4}} \\ \hline \hline
\begin{tabular}[c]{@{}l@{}}Sup. oracle MVDR~\cite{nakatani2017integrating} \end{tabular} &17.5      &5.2     &11.3      &13.0     &5.3   &9.2 \\
\begin{tabular}[c]{@{}l@{}}Sup. oracle GEV~\cite{heymann2016neural} \end{tabular} &17.0      &{{5.6}}     & {11.3}     &13.0     &{5.3}   & 9.2 \\ \hline 
\end{tabular}
\label{table:4}}
\vspace{-0.3cm}
\end{table}


\section{summary }\label{sec:summary}
In summary, we have proposed an unsupervised mask estimation approach for the GEV beamforming. The mask estimation is based on the joint estimation of the late reverberation component and a spatial filter that performs the beamforming to identify the clean source signal. This estimation is based on a maximum likelihood framework in a multi-channel linear prediction setting. The estimate of the clean source signal is used in the neural mask estimator to generate the speech presence probability which is in turn used in the generalized eigen value beamforming. Several ASR experiments on the CHiME-3 and the Reverb Challenge datasets confirm that the proposed approach of unsupervised mask estimation achieves performance similar to the supervised oracle mask estimation using paired clean and noisy audio recordings.  

\newpage


\bibliographystyle{IEEEbib}
\bibliography{mybib,refs}

\end{document}